%% file: draft6.tex
\documentstyle[12pt,psfig]{article}

\newcommand{\beq}{\begin{equation}}
\newcommand{\eeq}{\end{equation}}
\newcommand{\beqa}{\begin{eqnarray}}
\newcommand{\eeqa}{\end{eqnarray}}

\newcommand{\slas}[1]{{#1}\!\!/}
\newcommand{\slashK}{K\!\!\!\!/}
\newcommand{\etal}{{\em et al.}}

\newcommand{\PRD}[3]{Phys.\ Rev.\ D {\bf {#1}}, {#2} ({#3})}
\newcommand{\PRL}[3]{Phys.\ Rev.\ Lett.\ {\bf {#1}}, {#2} ({#3})}
\newcommand{\NPA}[3]{Nucl.\ Phys.\ A {\bf {#1}}, {#2} ({#3})}

\newcommand{\PLB}[3]{Phys.\ Lett.\ B {\bf {#1}}, {#2} ({#3})}
\newcommand{\ZPC}[3]{Z. Phys.\ C {\bf {#1}}, {#2} ({#3})}

\begin{document}

\begin{titlepage}
\renewcommand{\thefootnote}{\fnsymbol{footnote}}
\makebox[2cm]{}\\[-1in]
\begin{flushright}
TUM/T39-98-21
\end{flushright}
\vskip0.4cm
\begin{center}
  {\Large\bf
    \setcounter{footnote}{1}
    Exclusive $J/\psi$ photoproduction
    and gluon polarization\footnote{Work supported in part by BMBF}
  }\\ 

\vspace{2cm}

\setcounter{footnote}{0}
M. V\"anttinen\footnote{Alexander von Humboldt fellow}$^{,1}$
and L. Mankiewicz$^{1,2}$

\vspace{1 cm}

{\em $^1$ Physik Department, Technische Universit\"{a}t M\"{u}nchen, \\
D-85747 Garching, Germany}

\vspace{0.5 cm}

{\em $^2$ N. Copernicus Astronomical Center, Polish Academy of Science, \\
ul.\ Bartycka 18, PL--00-716 Warsaw}

\vspace{1cm}

{\em \today}

\vspace{1cm}

{\bf Abstract:\\[5pt]} \parbox[t]{\textwidth}{In exclusive
$J/\psi$ production by polarized photons incident on polarized
protons, a finite polarization asymmetry arises because of 
$c\bar c$ Fermi-motion and binding-energy effects. The asymmetry
depends on the polarized nonforward gluon distribution of the
proton and thus gives information on gluon polarization in the
proton. The analyzing power, however, is rather small.}

\end{center}
\end{titlepage}

\newpage
\renewcommand{\thefootnote}{\arabic{footnote}}
\setcounter{footnote}{0}


\setlength{\baselineskip}{8mm}

Gluon polarization in the proton is a topic of current interest
in hadronic physics (see e.g.\ \cite{COMPASS,RHIC}).
A useful probe of gluon distributions is the production of open or
hidden charm, because charm quarks dominantly couple to gluons and
not to light quarks in the proton. In this paper we discuss what the
exclusive photoproduction of $J/\psi$ mesons could tell about
the polarized gluon content of the proton.

In general, hard exclusive processes probe nonforward matrix elements
of the target nucleon, i.e.\ matrix elements of QCD operators between
nucleon states of different momenta \cite{Ji,Radyushkin}.
These matrix elements can be
expressed in terms of nonforward parton distributions, which are
functions of two momentum-fraction variables and generalize the usual
(forward) parton distributions. Nonforward distributions reduce to
forward distributions in the limit of equal proton momenta in the
initial and final states. This limit cannot be realized in exclusive
reactions because of simple kinematical reasons, but may be a good
approximation at high energy \cite{MPW97,Frankfurt,Martin97}.
Thus exclusive $J/\psi$
production in electron-proton collider experiments, e.g.\ at HERA,
could yield information on the small $x$ behaviour of the usual
gluon distribution of the proton, whereas fixed-target experiments
will probe the nonforward parton distributions in a more general case.

The dependence of unpolarized exclusive $J/\psi$ cross sections on
unpolarized gluon distributions has been discussed in Refs.\ 
\cite{Ryskin-unpolarized,Brodsky}. In principle, $J/\psi$ production
by polarized beams on polarized targets should depend on polarized
gluon distributions $\Delta G$. Unfortunately, contrary to earlier
belief, the production amplitude does not depend on $\Delta G$
in the first approximation (i.e.\ assuming a nonrelativistic $J/\psi$
bound state), as we showed in a recent paper \cite{MV-LM}.

However, as will be shown in detail in this paper, relativistic
corrections to $J/\psi$ production will depend on $\Delta G$.
Gluon polarization can therefore be accessed in $J/\psi$ production,
although the analyzing power may be rather small.

The expansion of $J/\psi$ amplitudes around the nonrelativistic
limit was first discussed by Keung and Muzinich \cite{KeungMuzinich}.
Their approach is based on expanding the perturbative amplitudes for
$c\bar c$ production in powers of the heavy quark relative momentum.
Higher Fock states of the $J/\psi$ are neglected, and the calculation
is therefore not gauge invariant. The issue of gauge invariance
was discussed more recently by Khan and Hoodbhoy \cite{KhanHoodbhoy}.
They pointed out that the gluonic contributions necessary to restore
gauge inariance are proportional to $v^3$, where $v$ is the relative
velocity of the charm quark and antiquark. Gluonic contributions are
thus subleading as compared to the first relativistic corrections in the
Keung--Muzinich approach, which are proportional to $v^2$.

We shall now calculate the $O(v^2)$ corrections to exclusive $J/\psi$
production following \cite{KeungMuzinich,KhanHoodbhoy}. We work in the
photoproduction limit (photon virtuality $q^2=-Q^2=0$) and consider
only the case of collinear scattering.

\begin{figure}
\centerline{\psfig{figure=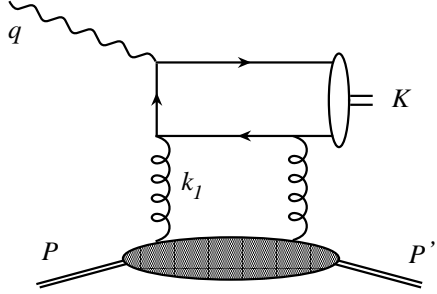,width=10cm}}
\caption{\sf One of 6 Feynman diagrams which contribute to the amplitude
(\protect\ref{partonic-amplitude1}). From Ref.\ \protect\cite{MV-LM}.} 
\label{fig:diagram}
\end{figure}

In terms of nonforward parton distributions $G$ and $\Delta G$, the
helicity amplitude for
\beq
  \gamma(q,\lambda) + p(p_1, S=\pm 1/2)
  \rightarrow J/\psi(K, \lambda'=\lambda) + p(p_2, S'=\pm 1/2)
\eeq
reads \cite{MV-LM}
\beqa
  {\cal A}_{\lambda\lambda \pm\pm}
  & = & \frac{1}{2} \int_{-1}^1 du 
        \frac{1}{(u-\xi+i\epsilon)(u+\xi-i\epsilon)} 
        \nonumber \\
  &   & \times \sum_{\lambda_1}
        \; A_{\lambda\lambda\lambda_1\lambda_1}
        \; \left[
        G(u,\xi;\mu^2) \pm \lambda_1 \, \Delta G(u,\xi;\mu^2)
        \right] \, ,
  \label{hadronic-amplitude-I}
\eeqa  
where $\xi = M^2/(2s-M^2)$ with $s=(q+p_1)^2$.        
The helicity amplitude $A_{\lambda\lambda'\lambda_1\lambda_2}$ for
\beq
 \gamma(q,\lambda) + g(k_1,\lambda_1)
 \rightarrow J/\psi(K,\lambda') + g(k_2,\lambda_2) \, ,
\eeq
where
\beq
  k_1 = \frac{u+\xi}{1+\xi} \; p_1 \, ,
\eeq
is the convolution of a perturbative $c\bar c$ production
amplitude\footnote{$H_{\lambda \lambda_1 \lambda_2}(q,K,k_1;\ell)$
denotes the $\gamma g \rightarrow c\bar c g$ amplitude with heavy-quark
spinors truncated, whereas in Ref.\ \protect\cite{MV-LM} we denoted
by $H^{\mu\nu}_{\lambda\lambda'}(q,K,k_1)$ the $\gamma g \rightarrow
J/\psi g$ amplitude with gluon polarization vectors truncated.} 
$H_{\lambda \lambda_1 \lambda_2}$ and a $J/\psi$ matrix element
(both are Dirac matrices):
\beqa
  \lefteqn{A_{\lambda\lambda'\lambda_1\lambda_2}(q,K,k_1)} \nonumber \\
  & = & {\rm Tr} \; \int \frac{d^4 \ell}{(2\pi)^4}
        H_{\lambda \lambda_1 \lambda_2}(q,K,k_1;\ell) 
        \int d^4x e^{i\ell\cdot x}
        \left\langle K,\lambda' \left| \psi(x/2)
        \bar\psi(-x/2) \right| 0 \right\rangle
        \, .
  \label{convolution}
\eeqa
There are six terms in $H_{\lambda \lambda_1 \lambda_2}$, corresponding
to permutations of the photon and the two gluons on the charm quark line.
A representative one, corresponding to the diagram in
Fig.~\ref{fig:diagram}, is
\beqa
  \lefteqn{H_{\lambda \lambda_1 \lambda_2}(q,K,k_1;\ell)} \nonumber \\
  & \sim & \slas{\epsilon}(q,\lambda) S_F(K/2-q+\ell)
           \slas{\epsilon}(k_1,\lambda_1) S_F(-K/2-k_2+\ell)
           \slas{\epsilon}^*(k_2,\lambda_2) \, ,
   \label{partonic-amplitude1}
\eeqa
where $S_F(p) \equiv (p\!\!\!/ - m_c)^{-1}$.
We have omitted normalization factors which cancel in a polarization
asymmetry calculation. Expanding the hard amplitude to second order in
$\ell$, we obtain
\beqa
  \lefteqn{A_{\lambda\lambda'\lambda_1\lambda_2}(q,K,k_1)} \nonumber \\
  & = & {\rm Tr} \, \Biggl[
        \left. H_{\lambda \lambda_1 \lambda_2}(q,K,k_1;\ell) \right|_{\ell=0}
        \left. \left\langle K,\lambda' \left| \psi(x/2) 
               \bar\psi(-x/2) \right| 0 \right\rangle  \right|_{x=0}
        \nonumber \\
  &   & \mbox{} + \left. \frac{\partial}{\partial\ell^\alpha}
        H_{\lambda \lambda_1 \lambda_2}(q,K,k_1;\ell) \right|_{\ell=0}
        \left. \left\langle K,\lambda' \left| \psi(x/2) 
               \left(
               -i\stackrel{\leftrightarrow}{\partial}_\alpha
               \right)
               \bar\psi(-x/2) \right| 0 \right\rangle  \right|_{x=0}
        \nonumber \\
  &   & \mbox{} + \left. \frac{1}{2}
        \frac{\partial^2}{\partial\ell^\alpha\partial\ell^\beta}
        H_{\lambda \lambda_1 \lambda_2}(q,K,k_1;\ell) \right|_{\ell=0}
        \nonumber \\
  &   & \times
        \left. \left\langle K,\lambda' \left| \psi(x/2) 
               \left(
               -i\stackrel{\leftrightarrow}{\partial}_\alpha
               \right)
               \left(
               -i\stackrel{\leftrightarrow}{\partial}_\beta
               \right)
               \bar\psi(-x/2) \right| 0 \right\rangle
        \right|_{x=0} \Biggr] \, .
 \label{expansion}
\eeqa
The $J/\psi$ matrix elements can be expressed as \cite{KhanHoodbhoy,Hoodbhoy}
\beqa
   \lefteqn{\left.\left\langle K,\lambda' \left| \psi(x/2) 
         \bar\psi(-x/2) \right| 0 \right\rangle  \right|_{x=0}}
         \nonumber \\
   & = & \frac{1}{2} M^{1/2} \left( \phi + \frac{\nabla^2\phi}{M^2} \right)
         \slas{\epsilon}^* \left( 1 + \frac{\slashK}{M} \right)
         - \frac{1}{6} M^{1/2} \frac{\nabla^2\phi}{M^2}
         \slas{\epsilon}^* \left( 1 - \frac{\slashK}{M} \right) \, ,
         \label{projector0} \\
   \lefteqn{\left.\left\langle K,\lambda' \left| \psi(x/2) 
         \left( -i \stackrel{\leftrightarrow}{\partial}_\alpha \right)
         \bar\psi(-x/2) \right| 0 \right\rangle  \right|_{x=0}}
         \nonumber \\
   & = & -\frac{1}{3} M^{3/2} \frac{\nabla^2\phi}{M^2}
         \epsilon^{*\beta} \left( g_{\alpha\beta}
         + i\epsilon_{\alpha\beta\mu\nu} \gamma^\mu \gamma_5 \frac{K^\nu}{M}
         \right) \, ,
         \label{projector1} \\
   \lefteqn{\left.\left\langle K,\lambda' \left| \psi(x/2) 
         \left( -i \stackrel{\leftrightarrow}{\partial}_\alpha \right)
         \left( -i \stackrel{\leftrightarrow}{\partial}_\beta  \right)
         \bar\psi(-x/2) \right| 0 \right\rangle  \right|_{x=0}}  
         \nonumber \\
   & = & \frac{1}{6} M^{5/2} \frac{\nabla^2\phi}{M^2}
         \left( g_{\alpha\beta} -  \frac{K_\alpha K_\beta}{M^2} \right)
         \slas{\epsilon}^* \left( 1 + \frac{\slashK}{M} \right) \, .
         \label{projector2}
\eeqa
The amplitude given by eqs.\ (\ref{expansion}-\ref{projector2}) is
gauge invariant at $O(v^2)$. In eqs.\ (\ref{projector0}-\ref{projector2}),
$\phi$ and $\nabla^2\phi$ are the $J/\psi$ meson's Coulomb-gauge wavefunction
and its Laplacian at the origin. For a wavefunction $\phi \sim e^{-r/a}$,
corresponding to a $1/r$ heavy-quark potential appropriate for
a heavy quarkonium state, $\nabla^2\phi$ is actually infinite at the origin.
The quantity $\nabla^2\phi$ in eqs.\ (\ref{projector0}-\ref{projector2})
has to be understood as representing the average of the Laplacian over a
region of volume $\sim 1/m_c^3$ and thus regarded as a free parameter.

Further corrections of the same magnitude arise because the quark mass
$m_c$ which appears in perturbative propagators is not exactly one half
of the physical charmonium mass which appears in the matrix elements
(\ref{projector0}-\ref{projector1}). These binding-energy corrections
are obtained by expanding the amplitude to first order in
$\epsilon_B = 2m_c-M$.

The parton-level amplitude simplifies to
\beqa
  \lefteqn{A_{\lambda\lambda'\lambda_1\lambda_2}(q,K,k_1)} \nonumber \\
  & \sim & \left( 1 - \frac{\epsilon_B}{2M}
                    + \frac{\nabla^2\phi}{3M^2\phi} \right)
           (\epsilon_1 \cdot \epsilon_2^*)
           (\epsilon_\gamma \cdot \epsilon_\psi^*)
           \nonumber \\
  &      & \mbox{} + \left( \frac{\epsilon_B}{2M}
                            - \frac{\nabla^2\phi}{M^2\phi} \right)
           \frac{M^2}{\hat s (\hat s - M^2)}
           \left[ M^2 (\epsilon_1 \cdot \epsilon_2^*)
                      (\epsilon_\gamma \cdot \epsilon_\psi^*)
           \right.
           \nonumber \\
  &      & \left. \mbox{} - \hat s (\epsilon_1 \cdot \epsilon_\psi^*)
                                   (\epsilon_\gamma \cdot \epsilon_2^*)
                  + (\hat s - M^2) (\epsilon_1 \cdot \epsilon_\gamma)
                                   (\epsilon_\psi^* \cdot \epsilon_2^*)
           \right] \, ,
\eeqa
where $\hat s = (q+k_1)^2$ and $\epsilon_1, \epsilon_2, \epsilon_\gamma,
\epsilon_\psi$ are the usual transverse polarization vectors for the
gluons, the photon and the $J/\psi$. Inserting these into the
hadron-level amplitude (\ref{hadronic-amplitude-I}) gives
\beqa
  {\cal A}_{\lambda\lambda \pm\pm}
  & \sim & \int_{-1}^1 du \left\{
           \left( \frac{1}{u-\xi+i\epsilon} - \frac{1}{u+\xi-i\epsilon} \right)
           \, G(u,\xi;\mu^2) [1 + O(\eta)] \right.
           \nonumber \\
  &      & \left. \mbox{} \pm \lambda \xi \frac{\eta}{2} \,
           \left( \frac{1}{(u-\xi+i\epsilon)^2}
                  - \frac{1}{(u+\xi-i\epsilon)^2} \right)
           \, \Delta G(u,\xi;\mu^2) \right\} \, ,
  \label{hadronic-amplitude-II}
\eeqa  
where
\beq
  \eta = \frac{\epsilon_B}{M} - 2\frac{\nabla^2\phi}{M^2\phi} \, .
\eeq

In order to estimate the magnitude of the asymmetry by using model
distributions, we shall employ the formalism of nonforward double
distributions $G(x,y)$ introduced by Radyushkin \cite{Radyushkin}.
Using
\beq
  G(u,\xi)
  = \int_0^1 dx \int_0^{1-x} dy \, G(x,y)
    \delta(u - [x + (x+2y-1)\xi])
\eeq
and a similar relation for the polarized distribution, we can 
express the beam-target polarization asymmetry as
\beq
  \frac{d\sigma(\uparrow\uparrow) - d\sigma(\uparrow\downarrow)}{
        d\sigma(\uparrow\uparrow) + d\sigma(\uparrow\downarrow)}
          \nonumber \\
  = \eta \; \frac{{\rm Re}\, (I_G^* I_{\Delta G})}{|I_G|^2}
\eeq
where
\beqa
  {\rm Re}\, I_G
    & = & 2 \int_0^1 dx \int_0^{1-x} dy
          \ln |(x+2y)^2 - (\bar\omega x)^2|
          \frac{\partial G(x,y)}{\partial y} \, ,
          \label{Re-G} \\
  {\rm Re}\, I_{\Delta G}
    & = & -\int_0^1 dx \int_0^{1-x} dy   
          \ln \left| \frac{x+2y-\bar\omega x}{x+2y+\bar\omega x} \right|
          \frac{\partial^2 \Delta G(x,y)}{\partial y^2} \, ,
          \label{Re-DeltaG} \\
  {\rm Im}\, I_G 
    & = & -2\pi \int_0^{2/(1+\bar\omega)} dx
          \left. G(x,y) \right|_{y = (\bar\omega-1)x/2} \, ,
          \label{Im-G} \\
  {\rm Im}\, I_{\Delta G}
    & = & \pi \int_0^{2/(1+\bar\omega)} dx
          \left. \frac{\partial \Delta G(x,y)}{\partial y}
          \right|_{y = (\bar\omega-1)x/2}
          \label{Im-DeltaG}
\eeqa
(we used the notation $\bar\omega = 1/\xi$). Thus the asymmetry is
proportional to an unknown parameter $\eta$ and depends on integrals
of both $G$ and $\Delta G$. Derivatives of the distributions
appear because we integrated by parts.

The values of the parameters $\epsilon_B/M$ and $\nabla^2\phi/(M^2\phi)$
were estimated in \cite{KhanHoodbhoy}, where it was reported that the
choice $\epsilon_B/M = -0.076$ and $\nabla^2\phi/(M^2\phi) = -0.073$
gives agreement with data on charmonium decays and inelastic $J/\psi$
photoproduction. In accordance with this, we shall use $\eta=0.07$ below.

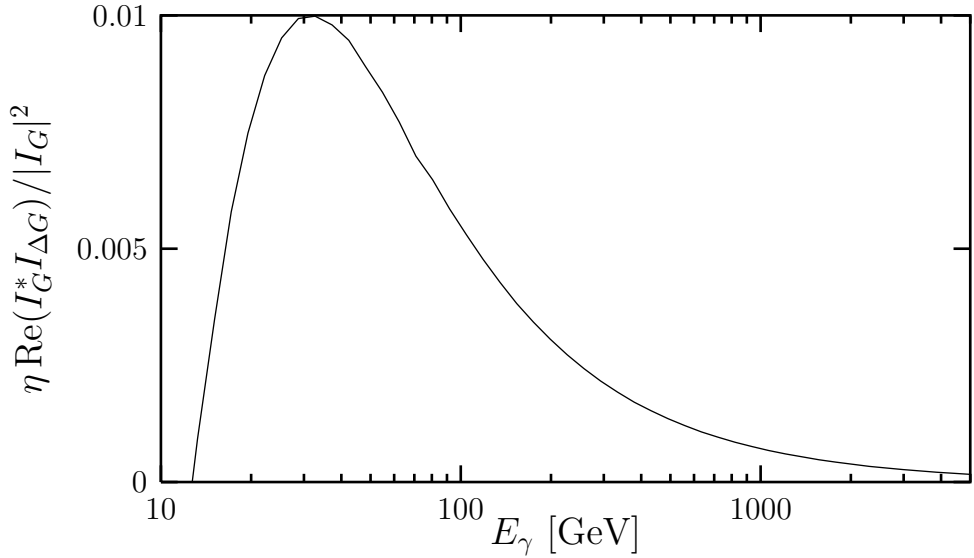
\begin{figure}
\input{fig2.tex}
\caption{\sf Polarization asymmetry obtained by using the model
distributions of eqs.\ 
(\protect\ref{model-unpolarized}--\protect\ref{model-polarized})
and setting $\eta=0.07$, plotted as a function of photon energy in
the proton rest frame.}
\label{fig:ratio}
\end{figure}

We now evaluate the integrals (\ref{Re-G}-\ref{Im-DeltaG}) for simple
model distributions and plot the resulting polarization asymmetry.
Following the discussion of Ref.\ \cite{Radyushkin-model}, we choose 
\beqa
  G(x,y)        & = & \frac{30}{(1-x)^5} \, [y(1-x-y)]^2 \, xg(x) \, ,
  \label{model-unpolarized} \\
  \Delta G(x,y) & = & \frac{30}{(1-x)^5} \, [y(1-x-y)]^2 \, x\Delta g(x) \, .
  \label{model-polarized}
\eeqa
We use the GRV-LO unpolarized gluon distribution $g(x)$ \cite{GRV-LO}
and the Gehr\-mann--Stirling A(LO) polarized distribution $\Delta g(x)$
\cite{Gehrmann}, evaluated at $Q^2 = 4 \, ({\rm GeV})^2$.
 Fig.~\ref{fig:ratio} shows the asymmetry
$\eta \, {\rm Re}(I_G^* I_{\Delta G})/|I_G|^2$ for $\eta=0.07$
as a function of photon energy in the proton rest frame. 
The asymmetry is of $O(10^{-2})$ in the photon energy range relevant
for fixed-target energies. At photon energies relevant for e.g.\ the
HERA collider experiments, the asymmetry drops by an order of magnitude. 

In summary, we have shown that a small but finite polarization asymmetry
arises from charm quark Fermi-motion and binding-energy corrections 
to polarized exclusive $J/\psi$ photoproduction. The asymmetry
depends on particular integrals of the polarized and unpolarized
nonforward parton distributions of the proton. Using a simple model
for parton distributions and previous estimates of $J/\psi$ parameters,
we estimate that the asymmetry may be of $O(10^{-2})$ at fixed-target
energies and $O(10^{-3})$ at collider energies.

\bigskip
 
{\bf Acknowledgement.} We wish to thank P.~Hoodbhoy and W.-D.~Nowak
for discussions.

\end{document}

%% file: fig2.tex
\setlength{\unitlength}{0.1bp}
\special{!
/gnudict 120 dict def
gnudict begin
/Color false def
/Solid false def
/gnulinewidth 5.000 def
/userlinewidth gnulinewidth def
/vshift -33 def
/dl {10 mul} def
/hpt_ 31.5 def
/vpt_ 31.5 def
/hpt hpt_ def
/vpt vpt_ def
/M {moveto} bind def
/L {lineto} bind def
/R {rmoveto} bind def
/V {rlineto} bind def
/vpt2 vpt 2 mul def
/hpt2 hpt 2 mul def
/Lshow { currentpoint stroke M
  0 vshift R show } def
/Rshow { currentpoint stroke M
  dup stringwidth pop neg vshift R show } def
/Cshow { currentpoint stroke M
  dup stringwidth pop -2 div vshift R show } def
/UP { dup vpt_ mul /vpt exch def hpt_ mul /hpt exch def
  /hpt2 hpt 2 mul def /vpt2 vpt 2 mul def } def
/DL { Color {setrgbcolor Solid {pop []} if 0 setdash }
 {pop pop pop Solid {pop []} if 0 setdash} ifelse } def
/BL { stroke gnulinewidth 2 mul setlinewidth } def
/AL { stroke gnulinewidth 2 div setlinewidth } def
/UL { gnulinewidth mul /userlinewidth exch def } def
/PL { stroke userlinewidth setlinewidth } def
/LTb { BL [] 0 0 0 DL } def
/LTa { AL [1 dl 2 dl] 0 setdash 0 0 0 setrgbcolor } def
/LT0 { PL [] 1 0 0 DL } def
/LT1 { PL [4 dl 2 dl] 0 1 0 DL } def
/LT2 { PL [2 dl 3 dl] 0 0 1 DL } def
/LT3 { PL [1 dl 1.5 dl] 1 0 1 DL } def
/LT4 { PL [5 dl 2 dl 1 dl 2 dl] 0 1 1 DL } def
/LT5 { PL [4 dl 3 dl 1 dl 3 dl] 1 1 0 DL } def
/LT6 { PL [2 dl 2 dl 2 dl 4 dl] 0 0 0 DL } def
/LT7 { PL [2 dl 2 dl 2 dl 2 dl 2 dl 4 dl] 1 0.3 0 DL } def
/LT8 { PL [2 dl 2 dl 2 dl 2 dl 2 dl 2 dl 2 dl 4 dl] 0.5 0.5 0.5 DL } def
/Pnt { stroke [] 0 setdash
   gsave 1 setlinecap M 0 0 V stroke grestore } def
/Dia { stroke [] 0 setdash 2 copy vpt add M
  hpt neg vpt neg V hpt vpt neg V
  hpt vpt V hpt neg vpt V closepath stroke
  Pnt } def
/Pls { stroke [] 0 setdash vpt sub M 0 vpt2 V
  currentpoint stroke M
  hpt neg vpt neg R hpt2 0 V stroke
  } def
/Box { stroke [] 0 setdash 2 copy exch hpt sub exch vpt add M
  0 vpt2 neg V hpt2 0 V 0 vpt2 V
  hpt2 neg 0 V closepath stroke
  Pnt } def
/Crs { stroke [] 0 setdash exch hpt sub exch vpt add M
  hpt2 vpt2 neg V currentpoint stroke M
  hpt2 neg 0 R hpt2 vpt2 V stroke } def
/TriU { stroke [] 0 setdash 2 copy vpt 1.12 mul add M
  hpt neg vpt -1.62 mul V
  hpt 2 mul 0 V
  hpt neg vpt 1.62 mul V closepath stroke
  Pnt  } def
/Star { 2 copy Pls Crs } def
/BoxF { stroke [] 0 setdash exch hpt sub exch vpt add M
  0 vpt2 neg V  hpt2 0 V  0 vpt2 V
  hpt2 neg 0 V  closepath fill } def
/TriUF { stroke [] 0 setdash vpt 1.12 mul add M
  hpt neg vpt -1.62 mul V
  hpt 2 mul 0 V
  hpt neg vpt 1.62 mul V closepath fill } def
/TriD { stroke [] 0 setdash 2 copy vpt 1.12 mul sub M
  hpt neg vpt 1.62 mul V
  hpt 2 mul 0 V
  hpt neg vpt -1.62 mul V closepath stroke
  Pnt  } def
/TriDF { stroke [] 0 setdash vpt 1.12 mul sub M
  hpt neg vpt 1.62 mul V
  hpt 2 mul 0 V
  hpt neg vpt -1.62 mul V closepath fill} def
/DiaF { stroke [] 0 setdash vpt add M
  hpt neg vpt neg V hpt vpt neg V
  hpt vpt V hpt neg vpt V closepath fill } def
/Pent { stroke [] 0 setdash 2 copy gsave
  translate 0 hpt M 4 {72 rotate 0 hpt L} repeat
  closepath stroke grestore Pnt } def
/PentF { stroke [] 0 setdash gsave
  translate 0 hpt M 4 {72 rotate 0 hpt L} repeat
  closepath fill grestore } def
/Circle { stroke [] 0 setdash 2 copy
  hpt 0 360 arc stroke Pnt } def
/CircleF { stroke [] 0 setdash hpt 0 360 arc fill } def
/C0 { BL [] 0 setdash 2 copy moveto vpt 90 450  arc } bind def
/C1 { BL [] 0 setdash 2 copy        moveto
       2 copy  vpt 0 90 arc closepath fill
               vpt 0 360 arc closepath } bind def
/C2 { BL [] 0 setdash 2 copy moveto
       2 copy  vpt 90 180 arc closepath fill
               vpt 0 360 arc closepath } bind def
/C3 { BL [] 0 setdash 2 copy moveto
       2 copy  vpt 0 180 arc closepath fill
               vpt 0 360 arc closepath } bind def
/C4 { BL [] 0 setdash 2 copy moveto
       2 copy  vpt 180 270 arc closepath fill
               vpt 0 360 arc closepath } bind def
/C5 { BL [] 0 setdash 2 copy moveto
       2 copy  vpt 0 90 arc
       2 copy moveto
       2 copy  vpt 180 270 arc closepath fill
               vpt 0 360 arc } bind def
/C6 { BL [] 0 setdash 2 copy moveto
      2 copy  vpt 90 270 arc closepath fill
              vpt 0 360 arc closepath } bind def
/C7 { BL [] 0 setdash 2 copy moveto
      2 copy  vpt 0 270 arc closepath fill
              vpt 0 360 arc closepath } bind def
/C8 { BL [] 0 setdash 2 copy moveto
      2 copy vpt 270 360 arc closepath fill
              vpt 0 360 arc closepath } bind def
/C9 { BL [] 0 setdash 2 copy moveto
      2 copy  vpt 270 450 arc closepath fill
              vpt 0 360 arc closepath } bind def
/C10 { BL [] 0 setdash 2 copy 2 copy moveto vpt 270 360 arc closepath fill
       2 copy moveto
       2 copy vpt 90 180 arc closepath fill
               vpt 0 360 arc closepath } bind def
/C11 { BL [] 0 setdash 2 copy moveto
       2 copy  vpt 0 180 arc closepath fill
       2 copy moveto
       2 copy  vpt 270 360 arc closepath fill
               vpt 0 360 arc closepath } bind def
/C12 { BL [] 0 setdash 2 copy moveto
       2 copy  vpt 180 360 arc closepath fill
               vpt 0 360 arc closepath } bind def
/C13 { BL [] 0 setdash  2 copy moveto
       2 copy  vpt 0 90 arc closepath fill
       2 copy moveto
       2 copy  vpt 180 360 arc closepath fill
               vpt 0 360 arc closepath } bind def
/C14 { BL [] 0 setdash 2 copy moveto
       2 copy  vpt 90 360 arc closepath fill
               vpt 0 360 arc } bind def
/C15 { BL [] 0 setdash 2 copy vpt 0 360 arc closepath fill
               vpt 0 360 arc closepath } bind def
/Rec   { newpath 4 2 roll moveto 1 index 0 rlineto 0 exch rlineto
       neg 0 rlineto closepath } bind def
/Square { dup Rec } bind def
/Bsquare { vpt sub exch vpt sub exch vpt2 Square } bind def
/S0 { BL [] 0 setdash 2 copy moveto 0 vpt rlineto BL Bsquare } bind def
/S1 { BL [] 0 setdash 2 copy vpt Square fill Bsquare } bind def
/S2 { BL [] 0 setdash 2 copy exch vpt sub exch vpt Square fill Bsquare } bind def
/S3 { BL [] 0 setdash 2 copy exch vpt sub exch vpt2 vpt Rec fill Bsquare } bind def
/S4 { BL [] 0 setdash 2 copy exch vpt sub exch vpt sub vpt Square fill Bsquare } bind def
/S5 { BL [] 0 setdash 2 copy 2 copy vpt Square fill
       exch vpt sub exch vpt sub vpt Square fill Bsquare } bind def
/S6 { BL [] 0 setdash 2 copy exch vpt sub exch vpt sub vpt vpt2 Rec fill Bsquare } bind def
/S7 { BL [] 0 setdash 2 copy exch vpt sub exch vpt sub vpt vpt2 Rec fill
       2 copy vpt Square fill
       Bsquare } bind def
/S8 { BL [] 0 setdash 2 copy vpt sub vpt Square fill Bsquare } bind def
/S9 { BL [] 0 setdash 2 copy vpt sub vpt vpt2 Rec fill Bsquare } bind def
/S10 { BL [] 0 setdash 2 copy vpt sub vpt Square fill 2 copy exch vpt sub exch vpt Square fill
       Bsquare } bind def
/S11 { BL [] 0 setdash 2 copy vpt sub vpt Square fill 2 copy exch vpt sub exch vpt2 vpt Rec fill
       Bsquare } bind def
/S12 { BL [] 0 setdash 2 copy exch vpt sub exch vpt sub vpt2 vpt Rec fill Bsquare } bind def
/S13 { BL [] 0 setdash 2 copy exch vpt sub exch vpt sub vpt2 vpt Rec fill
       2 copy vpt Square fill Bsquare } bind def
/S14 { BL [] 0 setdash 2 copy exch vpt sub exch vpt sub vpt2 vpt Rec fill
       2 copy exch vpt sub exch vpt Square fill Bsquare } bind def
/S15 { BL [] 0 setdash 2 copy Bsquare fill Bsquare } bind def
/D0 { gsave translate 45 rotate 0 0 S0 stroke grestore } bind def
/D1 { gsave translate 45 rotate 0 0 S1 stroke grestore } bind def
/D2 { gsave translate 45 rotate 0 0 S2 stroke grestore } bind def
/D3 { gsave translate 45 rotate 0 0 S3 stroke grestore } bind def
/D4 { gsave translate 45 rotate 0 0 S4 stroke grestore } bind def
/D5 { gsave translate 45 rotate 0 0 S5 stroke grestore } bind def
/D6 { gsave translate 45 rotate 0 0 S6 stroke grestore } bind def
/D7 { gsave translate 45 rotate 0 0 S7 stroke grestore } bind def
/D8 { gsave translate 45 rotate 0 0 S8 stroke grestore } bind def
/D9 { gsave translate 45 rotate 0 0 S9 stroke grestore } bind def
/D10 { gsave translate 45 rotate 0 0 S10 stroke grestore } bind def
/D11 { gsave translate 45 rotate 0 0 S11 stroke grestore } bind def
/D12 { gsave translate 45 rotate 0 0 S12 stroke grestore } bind def
/D13 { gsave translate 45 rotate 0 0 S13 stroke grestore } bind def
/D14 { gsave translate 45 rotate 0 0 S14 stroke grestore } bind def
/D15 { gsave translate 45 rotate 0 0 S15 stroke grestore } bind def
/DiaE { stroke [] 0 setdash vpt add M
  hpt neg vpt neg V hpt vpt neg V
  hpt vpt V hpt neg vpt V closepath stroke } def
/BoxE { stroke [] 0 setdash exch hpt sub exch vpt add M
  0 vpt2 neg V hpt2 0 V 0 vpt2 V
  hpt2 neg 0 V closepath stroke } def
/TriUE { stroke [] 0 setdash vpt 1.12 mul add M
  hpt neg vpt -1.62 mul V
  hpt 2 mul 0 V
  hpt neg vpt 1.62 mul V closepath stroke } def
/TriDE { stroke [] 0 setdash vpt 1.12 mul sub M
  hpt neg vpt 1.62 mul V
  hpt 2 mul 0 V
  hpt neg vpt -1.62 mul V closepath stroke } def
/PentE { stroke [] 0 setdash gsave
  translate 0 hpt M 4 {72 rotate 0 hpt L} repeat
  closepath stroke grestore } def
/CircE { stroke [] 0 setdash 
  hpt 0 360 arc stroke } def
/Opaque { gsave closepath 1 setgray fill grestore 0 setgray closepath } def
/DiaW { stroke [] 0 setdash vpt add M
  hpt neg vpt neg V hpt vpt neg V
  hpt vpt V hpt neg vpt V Opaque stroke } def
/BoxW { stroke [] 0 setdash exch hpt sub exch vpt add M
  0 vpt2 neg V hpt2 0 V 0 vpt2 V
  hpt2 neg 0 V Opaque stroke } def
/TriUW { stroke [] 0 setdash vpt 1.12 mul add M
  hpt neg vpt -1.62 mul V
  hpt 2 mul 0 V
  hpt neg vpt 1.62 mul V Opaque stroke } def
/TriDW { stroke [] 0 setdash vpt 1.12 mul sub M
  hpt neg vpt 1.62 mul V
  hpt 2 mul 0 V
  hpt neg vpt -1.62 mul V Opaque stroke } def
/PentW { stroke [] 0 setdash gsave
  translate 0 hpt M 4 {72 rotate 0 hpt L} repeat
  Opaque stroke grestore } def
/CircW { stroke [] 0 setdash 
  hpt 0 360 arc Opaque stroke } def
/BoxFill { gsave Rec 1 setgray fill grestore } def
end
}
\begin{picture}(3600,2160)(0,0)
\special{"
gnudict begin
gsave
0 0 translate
0.100 0.100 scale
0 setgray
newpath
1.000 UL
LTb
450 300 M
63 0 V
2987 0 R
-63 0 V
450 1180 M
63 0 V
2987 0 R
-63 0 V
450 2060 M
63 0 V
2987 0 R
-63 0 V
450 300 M
0 63 V
0 1697 R
0 -63 V
790 300 M
0 31 V
0 1729 R
0 -31 V
989 300 M
0 31 V
0 1729 R
0 -31 V
1130 300 M
0 31 V
0 1729 R
0 -31 V
1240 300 M
0 31 V
0 1729 R
0 -31 V
1329 300 M
0 31 V
0 1729 R
0 -31 V
1405 300 M
0 31 V
0 1729 R
0 -31 V
1471 300 M
0 31 V
0 1729 R
0 -31 V
1528 300 M
0 31 V
0 1729 R
0 -31 V
1580 300 M
0 63 V
0 1697 R
0 -63 V
1920 300 M
0 31 V
0 1729 R
0 -31 V
2119 300 M
0 31 V
0 1729 R
0 -31 V
2260 300 M
0 31 V
0 1729 R
0 -31 V
2370 300 M
0 31 V
0 1729 R
0 -31 V
2459 300 M
0 31 V
0 1729 R
0 -31 V
2535 300 M
0 31 V
0 1729 R
0 -31 V
2601 300 M
0 31 V
0 1729 R
0 -31 V
2658 300 M
0 31 V
0 1729 R
0 -31 V
2710 300 M
0 63 V
0 1697 R
0 -63 V
3050 300 M
0 31 V
0 1729 R
0 -31 V
3249 300 M
0 31 V
0 1729 R
0 -31 V
3390 300 M
0 31 V
0 1729 R
0 -31 V
3500 300 M
0 31 V
0 1729 R
0 -31 V
1.000 UL
LTb
450 300 M
3050 0 V
0 1760 V
-3050 0 V
450 300 L
1.000 UL
LT0
568 300 M
20 163 V
64 449 V
63 407 V
63 297 V
63 217 V
64 142 V
63 73 V
63 9 V
64 -33 V
63 -58 V
63 -99 V
64 -97 V
63 -112 V
63 -129 V
63 -89 V
64 -110 V
63 -97 V
63 -93 V
64 -87 V
63 -80 V
63 -70 V
63 -64 V
64 -59 V
63 -52 V
63 -47 V
64 -42 V
63 -38 V
63 -32 V
64 -30 V
63 -26 V
63 -24 V
63 -20 V
64 -19 V
63 -16 V
63 -15 V
64 -13 V
63 -11 V
63 -11 V
64 -9 V
63 -8 V
63 -8 V
63 -6 V
64 -6 V
63 -5 V
63 -5 V
64 -4 V
63 -4 V
stroke
grestore
end
showpage
}
\put(1975,100){\makebox(0,0){{\large $E_\gamma$ [GeV]}}}
\put(0,1180){%
\special{ps: gsave currentpoint currentpoint translate
270 rotate neg exch neg exch translate}%
\makebox(0,0)[b]{\shortstack{{\large $\eta \, {\rm Re}(I_G^* I_{\Delta G})/|I_G|^2$}}}%
\special{ps: currentpoint grestore moveto}%
}
\put(2710,200){\makebox(0,0){1000}}
\put(1580,200){\makebox(0,0){100}}
\put(450,200){\makebox(0,0){10}}
\put(400,2060){\makebox(0,0)[r]{0.01}}
\put(400,1180){\makebox(0,0)[r]{0.005}}
\put(400,300){\makebox(0,0)[r]{0}}
\end{picture}